\documentclass[preprint]{aastex62}
\usepackage{xcolor, soul}
\sethlcolor{green}
\hypersetup{linkcolor=red,citecolor=blue,filecolor=cyan,urlcolor=magenta}

\received{Month day, Year}
\revised{Month day, Year}
\accepted{Month day, Year}

\accepted{ApJ}

\shorttitle{On the Nature of Propagating Intensity Disturbances in Polar Plumes}
\shortauthors{Cho et al.}

\begin{document}

\title{On the Nature of Propagating Intensity Disturbances in Polar Plumes during the 2017 Total Solar Eclipse}
\correspondingauthor{Kyung-Suk Cho}
\email{kscho@kasi.re.kr}

\author[0000-0003-2161-9606]{Kyung-Suk Cho}
\affiliation{Space Science Division, Korea Astronomy and Space Science Institute, Daejeon 305-348, Republic of Korea}
\affiliation{Department of Astronomy and Space Science, University of Science and Technology, Daejeon 305-348, Republic of Korea}

\author{Il-Hyun Cho}
\affiliation{School of Space Research, Kyung Hee University, Yongin, 17104, Republic of Korea}

\author{Maria S. Madjarska}
\affiliation{Astronomy Program, Department of Physics and Astronomy, Seoul National University, Seoul 151-742, Republic of Korea}
\affiliation{Max Planck Institute for Solar System Research, Justus-von-Liebig-Wig 3, 37077, G$\ddot{o}$ttingen, Germany}

\author{Valery M. Nakariakov}
\affiliation{School of Space Research, Kyung Hee University, Yongin, 17104, Republic of Korea}
\affiliation{University of Warwick, Coventry, CV4 7AL, United Kingdom}

\author{Heesu Yang}
\affiliation{Space Science Division, Korea Astronomy and Space Science Institute, Daejeon 305-348, Republic of Korea}

\author{Seonghwan Choi}
\affiliation{Space Science Division, Korea Astronomy and Space Science Institute, Daejeon 305-348, Republic of Korea}

\author{Eun-Kyung Lim}
\affiliation{Space Science Division, Korea Astronomy and Space Science Institute, Daejeon 305-348, Republic of Korea}

\author{Kyung-Sun Lee}
\affiliation{Astronomy Program, Department of Physics and Astronomy, Seoul National University, Seoul 151-742, Republic of Korea}

\author{Jung-Jun Seough}
\affiliation{Space Science Division, Korea Astronomy and Space Science Institute, Daejeon 305-348, Republic of Korea}

\author{Jaeok Lee}
\affiliation{Space Science Division, Korea Astronomy and Space Science Institute, Daejeon 305-348, Republic of Korea}

\author{Yeon-Han Kim}
\affiliation{Space Science Division, Korea Astronomy and Space Science Institute, Daejeon 305-348, Republic of Korea}

\begin{abstract}
The propagating intensity disturbances (PIDs) in plumes are still poorly  understood and their identity (magnetoacoustic waves or flows) remains an open question. We investigate PIDs in five plumes located in the northern polar coronal hole observed during the 2017 total solar eclipse. Three plumes are associated with coronal bright points, jets and macrospicules at their base (active plumes) and the other two plumes are not (quiet plumes). The electron temperature at the base of the plumes is obtained from the filter ratio of images taken with the X-ray Telescope on board Hinode and the passband ratio around 400 nm from an eclipse instrument, the Diagnostic Coronagraph Experiment (DICE). The phase speed ($v_{r}$), frequency ($\omega$), and wavenumber ($k$) of the PIDs in the plumes are obtained by applying a Fourier transformation to the space-time ($r-t$ plane) plots in images taken with the Atmospheric Imaging Assembly (AIA) in three different wavelength channels (171 \AA, 193 \AA, and 211 \AA). We found that the PIDs in the higher temperature AIA channels, 193 and 211~\AA,  are faster than that of the cooler AIA 171~\AA\ channel. This tendency is more significant for the active plumes than the quiet ones. The observed speed ratio ($\sim$1.3) between the AIA 171 and 193~\AA\ channels is similar to the theoretical value (1.25) of a slow magnetoacoustic wave. Our results support the idea that PIDs in plumes represent a superposition of slow magnetoacoustic waves and plasma outflows that consist of dense cool flows and hot coronal jets.

\end{abstract}

\section{Introduction} \label{sec:intro}

Polar plumes are long-lasting, dense, ray-like, and relatively bright solar corona structures, usually observed in the  EUV, X-ray, and white light. They are magnetically open linear structures extending from the Sun's surface to about 30 R$_\odot$ and  are known to be a source of the fast solar wind \citep[][and the references therein]{2015LRSP...12....7P,2020arXiv200900627Z}.
\citet{2008ApJ...682L.137R} suggested that low-rate magnetic reconnection is responsible for the formation and sustainability of plumes. Plumes are thought to be associated with coronal bright points \citep[CBPs, ][]{2019LRSP...16....2M} that form at the boundaries of supergranulation cells. Several observations  support that polar plumes are associated with CBPs, X-ray jets, or EUV jets in the network regions. \citet{1995ApJ...452..457W} suggested that the properties of a few hours-lasting and hazy plume structure can be adequately explained by the quasi-static upward evaporation of material heated by conduction during magnetic reconnection at the boundary of the super-granular cells. This suggestion was supported by  the observations of a particularly bright plume taken with the Solar Heliospheric Observatory/Coronal Diagnostic Spectrometer (SoHO/CDS). \citet{1999A&A...350..286Y} investigated  a plume above the limb and the plume base visible on the solar disk. They reported that the temperature of the plume above the limb is about 1 MK, and the temperature remains constant with height. They found a strong brightening lying directly below the plume's main body with a high temperature of 2 MK , i.e. a CBP. They suggested that the footpoint morphology is similar to an emerging bipole in a region of unipolar magnetic flux which is typical for CBPs \citep{2019LRSP...16....2M}.

Due to the open magnetic configuration in polar coronal holes, polar plumes could be waveguides for magnetohydrodynamic (MHD) waves \citep[e.g.][]{2006RSPTA.364..473N, 2015LRSP...12....7P}. The existence of waves in plumes has been demonstrated by observations taken with UltraViolet Coronagraph Spectrometer \citep[UVCS, ][]{1997ApJ...491L.111O}, Extreme ultraviolet Imaging Telescope \citep[EIT, ][]{1998ApJ...501L.217D}, CDS \citep{2000SoPh..196...63B}, Solar Ultraviolet Measurements of Emitted Radiation \citep[SUMER, ][]{2012A&A...546A..93G} on board the SoHO, and by a theoretical model for the propagation of magnetoacoustic  waves in polar plumes by \citet{1999ApJ...514..441O}.

Propagating intensity disturbances (PIDs) in plumes have been interpreted as compressible, magnetoacoustic waves as their intensity enhancement coincides with the blue-shift of the O~{\sc v} 629~\AA\ line profile taken with CDS \citep{2000SoPh..196...63B}. \citet{2009A&A...499L..29B} observed intensity oscillations with a period of 10 to 30 min and propagation velocity from 75 to 125 km s$^{-1}$, depending on the line formation temperature. They concluded that the most likely cause for PIDs is slow magnetoacoustic waves. This conclusion was supported by \citet{2012A&A...546A..93G} who reported a detection of simultaneous intensity enhancement and Doppler velocities of PIDs. The observational evidence which favours the interpretation of PIDs in terms of slow magnetoacoustic waves is summarised in \citet{2011SSRv..158..267B}. Meanwhile, \citet{2010A&A...510L...2M} suggested  that plumes are associated with high speed jets which are multi-thermal and repeat quasi-periodically. Later, \citet{2011ApJ...736..130T} supported this by showing the significant asymmetries of spectral lines observed with the Extreme-ultraviolet Imaging Spectrometer/Hinode. On the other hand, \citet{2010ApJ...724L.194V} demonstrated that the line asymmetry is fully consistent with the interpretation of PIDs in terms of upward propagating compressive waves. \citet{2014ApJ...793...86P} concluded that PIDs represent plasma outflows because the expected temperature dependence of the magnetoacoustic wave was not observed. In their data, the apparent speed  (30 -- 300 km s$^{-1}$) is higher in the low temperature channel AIA 171~\AA\ than those in the high temperature channels AIA~193 and 211~\AA. On the other hand, \citet{2011A&A...528L...4K} established the temperature dependence of the apparent phase speed, theoretically predicted for slow magnetoacoustic waves. Moreover, \citet{2014A&A...568A..96G} found that the dissipation lengths of PIDs are frequency dependent, which is consistent with the slow wave behaviour. A recent analysis of damping and power spectra of PIDs in coronal holes, observed with SDO/AIA, link this phenomenon with slow magnetoacoustic waves too \citep{2016RAA....16...93J}. In addition, similar wave motions are routinely detected in the legs of coronal loops, and their observed properties are well described by the MHD wave theory \citep[e.g.][]{2009SSRv..149...65D, 2012RSPTA.370.3193D}.

Due to the overlapping of the plume structures with fore- and background emission and the lack of observations with a sufficiently high spatial resolution, the polar plume's triggering mechanism is still not well understood. For this reason, the observations of the footpoints of the polar plumes are vital as they can give a clear connection from the bipole emergence through the H$\alpha$/EUV/X-ray jets to the polar plumes. In this study, we investigate plumes in a polar coronal hole to inspect PIDs and their relationship with activities at the base of the plumes.
For this, we analyze polar plumes that are found in an image taken at the 2017 total solar eclipse by using the Diagnostic Coronagraph experiment (DICE), Hinode/XRT, SDO/AIA, and SoHO/LASCO images.
We divide the plumes into active and quiet ones, and derive the wavenumber and frequency of PIDs by using the Fourier component analysis to investigate a phase speed dependency on the temperature.

The following Section describes observation and data processing. In Section 3, we present space-time plots of plumes and dispersion relations of the propagation in the plumes. Finally, we summarize and discuss our results in Section~4.

\section{Observations and Data Processing} \label{sec:data}

The observation targets are plumes observed at the north polar limb during the total solar eclipse (TSE) on 2017 August 21 at a nearby Jackson hole, Wyoming. The obtained data from the eclipse observation are white light (WL) images and narrow-band polarization images around 400~nm. The WL images were taken using a simple camera system consisting of a Canon 400 mm lens with a 2X extender. During the TSE, we took WL images with different exposure times from 0.15 ms to 0.25 s to acquire features from the chromosphere to the outer corona ( $>$ 3  R$_\odot$).
After the image alignment and rotation, we applied the Multi-scale Gaussian Normalization method \citep{2014SoPh..289.2945M} to the WL images. This processing dramatically enhances the faint plume intensity signal at a high altitude. The pixel resolution of the WL images is about 1.03 arcsec.
Narrow-band images were taken using the Diagnostic Coronagraph experiment (DICE) consisting of an optical assembly, a filter wheel, a polarizer unit, and a CCD camera. Since the temperature determines the shape of the coronal spectrum (Thompson-scattered photospheric light), we used the polarization intensity ratio at the temperature-sensitive wavelengths at 393.9 and 402.5 nm, and determined the coronal temperature by applying the method suggested by \citet{1976SoPh...48....3C}.
The pixel size of the polarization data is about 12.7 arcsec, and we used 3 by 3 data binning of stacked images after aligning the images for specific filters and polarization angles to obtain a temperature map.
The temperature measurements using the pass-band ratio imaging has been conducted during the TSE \citep{2017JGRA..122.5856R} and Balloon mission \citep{2020Solar}.
More detailed description of the DICE instrument and data reduction can be found in \citet{2020JKAS...53...87C}.
First from the DICE observations we identified regions of enhanced emission associated with CBPs and their temperature was first determined.
To verify the CBPs and the temperature at the base of the plumes, we also checked  data taken with the X-ray telescope \citep[XRT,][]{2007SoPh..243...63G} on board Hinode \citep{2007SoPh..243....3K} during the eclipse time.
It was noted that the temperature structures in the DICE data are generally similar to those in the XRT images but CBPs could not be seen in the DICE data due to occulting the solar disk by the moon.
We used level 1 data of XRT processed by xrt\_prep.pro. The pixel resolution of the XRT images is about 1 arcsec.
By taking a full-Sun image observed at 16:43 UT, we calculated  the electron temperature by applying the xrt\_teem.pro to the `Al-poly' and `Al-mesh' filter data pair, which is sensitive for temperatures from Log T = 5.5 up to Log T = 6.3 \citep{2014SoPh..289.1029N}.

We also used SoHO \citep{1995SoPh..162....1D}/LASCO C2 \citep{1995SoPh..162..357B} and SDO \citep{2012SoPh..275....3P} /AIA \citep{2012SoPh..275...17L} data to inspect the off-limb plume structure and dynamics.
The level 0.5 data of LASCO C2 from 16:00 UT to 20:00 UT were used to produce an imaging data cube.
We normalized the data set in a temporal direction by subtracting a background image, including the F coronal and the instrument stray light.
We calculated all pixel median values between 2.2 R$_\odot$ and 5.8 R$_\odot$ in a temporal direction. We subtracted this value from the data set and divided by the corresponding standard absolute deviation.
We took an averaged image from the processed data cube after removing the blinking due to frame-to-frame variation in exposure time and spikes (stars) in the data.
The AIA 171, 193, 211, and 304~\AA\ data with a 12 s time cadence, and a {0.6$\arcsec$ per pixel scale were used to inspect the dynamics of the plume footpoints within the range from 1 to 1.26 R$_\odot$.
To produce  a space-time image $I(r, t)$,  we defined slit cuts along the plume lengths in the AIA data.
The position of a slit was determined by a linear fitting from the visually determined locations.
Thus, the inclined distance of the slits represents the distance from the solar surface. The intensities along the slits, and  time are used to investigate the plume dynamics. For a given distance, we divided the original time series by its temporal average and took the first order time-difference that is obtained by subtracting the previous intensity from the current one. Then we applied a median filter with the 3 minutes by 1.5 Mm to eliminate the noise.
In case of AIA 304~\AA, the space-time image is subtracted and then divided by the temporal mean and standard deviation to remove the background \citep{2006SoPh..236..263M}.

From the AIA slit images, we derived the properties of localized moving features in the plumes using Fourier transformation.
To calculate the propagation speed, we applied the Fourier transformation to the space-time image. The propagating feature in $I(r, t)$ appears in the Fourier power ($\vert F(k_r, w) \vert ^2$), and its phase speed could be defined as ($-\omega$/$k_r$). Hence, a positively propagating feature in $I(r, t)$ appears as a prominent power in the second and forth quadrants in the Fourier space \citep[e.g., Appendix in][]{2014ApJ...787..124D}. We set a slit along the plumes in the AIA channels to track the PIDs and generate an $I(r, t)$ image that contains both the temporal and spatial information of the propagation.
A Fourier transformation is applied  to the $I(r, t)$ image to obtain  its conjugate ($k_r$, $\omega$) plane.
This results in the observational dispersion relationship for the frequency ($\omega$) as a function of wavenumber ($k_r$).
Since moving features with a speed in the $I(r, t)$ plane are transformed to features lying along a diagonal line in the ($k_r$, $\omega$),
we can derive a particular speed of the propagation features by measuring the slope ($\omega$/$k_r$) of the diagonal line \citep[e.g., Figure~1 in][]{2014ApJ...787..124D}.

\section{Results} \label{sec:data}

For tracing the plumes to the solar surface, we used the SDO/AIA, SOHO/LASCO C2, and the eclipse data.
Figure 1 shows the WL eclipse image that is composed of SDO/AIA 171~\AA\ in the low corona and LASCO C2 images in the upper corona.
The polar coronal hole in the northern hemisphere contains dense and relatively bright plumes extending from the solar surface to about 5.5 R$_\odot$.
We used the eclipse image that fills a gap between the low corona ($<$ 1.2 R$_\odot$) and the upper corona (2.5 R$_\odot$ $\leq$ R $\leq$ 5.5 R$_\odot$).
In this study, we selected the plumes whose connection from the higher corona to the surface is clear and then marked them with arrows and labels as shown in Figure~1.
It is noted from the eclipse and LASCO images that the plumes (L1, C, and R1) in the middle are brighter and more extended into the higher corona than the others (L2 and R2). The plumes, `L1', `C', and `R1' correspond to the coronal fine structures, `3', `2', and `1', respectively, in the Figure 1 of \citet{2018ApJ...860..142H}.

We inspected the location where these plumes are rooted in the AIA images to study the physical properties of the plasma at the  plume footpoints.
Figure 2 shows the averaged intensity maps of AIA 171, 193, 211, and 304~\AA\ channels, from top to bottom respectively.
These maps are obtained from the AIA images with 12 seconds cadence during 100 minutes from 16:00 UT to 17:40 UT.
Except for the AIA 304~\AA\ channel, plume structures are seen in all AIA channels.
The plumes in the AIA 171~\AA\ channel \citep[maximum response temperature, T$_e$ $\sim$0.7 $\times$ 10$^6$ K from the Fe~{\sc ix} line,][]{2010A&A...521A..21O} are brighter and denser than those seen in the AIA 191 and 211~\AA\ channels.
The plumes are also detected in the high-temperature channels of AIA 193 ( 0.7 $\times$ 10$^6$ $\leq$ T$_e$ $\leq$ 1.6 $\times$ 10$^6$ K)
and AIA 211~\AA\ (T$_e$ $\sim$1.1 $\times$ 10$^6$ K and 1.4 $\times$ 10$^6$ K for  Fe~{\sc x} and  Fe~{\sc xi} lines, respectively).
In this study, we will refer to the high temperature channels AIA 193 and 211~\AA\ as AIAH and to the low temperature channel  the AIA 171~\AA\ as AIAC.
As it can be  noted  from Figure~2, cool and hot plasma co-exist in plumes.
 From the animation of Figure 2, we identified apparent outflows in the plumes in AIAH and AIAC, and recurrent spicules and macro-spicules in the base of the plumes from the 304~\AA\ channel animation.
In AIAH, EUV jets are identified in the plumes `L1', `C', and `R1' which have CBPs in their footpoints, and extend to the higher corona with brighter structures  seen in the LASCO C2 FOV as shown in Figure 1.

Based on the multi-site WL observations during the 2017 TSE, \citet{2018ApJ...860..142H} reported that these plumes are associated with the coronal (eclipse) jets,
while the L2 plume is not accompanied by WL jets seen during the eclipse. According to their study, the jets associated with the plumes were upwardly ejected with an apparent speed of about 450 km s$^{-1}$ beyond 2 R$_\odot$.
The R2 plume that has no jets, macrospicules, and a CBP, is not studied in their work.

As suggested by \citet{1995ApJ...452..457W}, the plumes' hazy structures in high-temperature imaging channels may be due to quasi-static upward evaporation heated by thermal conduction following magnetic reconnection.
This suggestion led us to inspect the temperature structure at the base of the plumes. The temperature was estimated from the Hinode/XRT and the DICE data by applying the methods described in the previous section.
Figure 3 shows EUV and soft X-ray intensity images and temperature maps from XRT and DICE.
The XRT image was taken with the `Al-poly' filter. It shows that EUV brightenings at the base of the plumes (L1, C, R1), are consistent with CBPs, and the temperature structures from XRT and DICE look similar to each other.
It reveals that the plumes (L1, C, R1) that have jets are associated with CBPs in their bases.
Their temperature ranges between 1 MK and 1.5 MK (Figure 3 (c) and (d)).
The other plumes, L2 and R2, have no CBPs at their bases, and their temperatures are not significantly high.

Flows of hot and cool plasmas in the plumes were investigated by using space-time plots along the plumes tracing them from the limb. By using the method described in Section 2, we produced space-time plots over the dashed bars in Figure 2 as shown in Figure 4.
In Figure 4, the oblique alternating bright and dark stripes represent the intensity disturbances that propagate upwardly along the plumes. The strips in AIAH look similar, and are more prominent in AIA 193~\AA\ due to a lower signal-to-noise of AIA 211~\AA. Most of the strips in AIAC coincide with those in AIAH. However,
we find that some faint AIAC strips between 17:00 UT and 17:10 UT are missing in AIAH as denoted by arrows on the panel of L1 plume in Figure 4, and vice versa (e.g., the jet in AIAH at 17:00 UT as marked by an arrow on the panel of the R1 plume).
In the jets associated plumes (L1, C, and R1), we find a macro-spicules that rise to up to 50\arcsec\ and later fall back (AIA 304 panels of Figure 4). There is a difference between the jet and the macrospicule activities.
For instance, the bright jet (black dashed line) in the C plume is associated with a weak macrospicule, while the big macrospicule at $\sim$17:00 UT is associated with a weak PID in AIAH and AIAC.
The one-to-one connection between the macro-spicules and the jets has ambiguity. It is quite clear for the L1 and R1 plumes, but not for the C plume.

Macro-spicules in Plume C ($\sim$17:10~UT) and L1 ($\sim$17:20~UT)  coincide with  jets.
However, the jet at $\sim$17:00~UT in the plume R1 occurs after the rising of the macro-spicule ($\sim$16:20~UT).
It is likely that the macrospicules are related to the CBPs in general.

As marked with the black dashed line in Figure 4, the inclination of a strip for the time axis can be used to estimate the individual strip speed.
Since we are interested in statistics of the PID speed at the different channels of AIA, we performed a Fourier transformation to the space-time data
and obtained a dispersion relation, as shown in Figure 5.
The Fourier transformation localizes moving features, and the features with speed in the space-time $I(r, t)$ plane are transformed to features lying along the diagonal line in the dispersion ($k_{r}$, $\omega$) plane.
Non-moving and quasi-stationary features exist at low $\omega$, and the noise sources are distributed isotropically.

To obtain the PID general properties, we obtained the mean Fourier powers of the intensity disturbances in all the plumes for the different AIA channels.
Figure 6 shows the dispersion relationships of all plumes in the various filters, in which the power is enhanced, and the noise is reduced.
To obtain an average speed of propagating packets with different wave number and frequency,
we set a diagonal dashed line laid along the features as shown in Figure 6, and obtained the propagation speed ($v_{r}$ = $-\omega$ / $k_{r}$) from the slop of the diagonal line.
When we visually set the line that passes through zero $\omega$ and $k_{r}$, we did not use the peaks of the power of plumes located at a small wave number and low frequency because they can be related to the jets.
In other words, we determined the velocity of the PIDs by considering the features in the high frequency (10 -- 20~min periods) rather than low frequency (20 -- 30 min periods) as in the previous reports by \citet{1998ApJ...501L.217D} and \citet{2000AIPC..537..160B}.
 For example, $\sim$166 km s$^{-1}$ in AIA 193 is calculated from wavenumber of -0.01 Mm$^{-1}$ (wavelength of 100~Mm) and frequency of 0.1 min$^{-1}$ (period of 10 min) as shown in the upper right panel of Figure 6.
For the different AIA channels, the disturbance speeds ($v_{r}$) are not the same but increase as the temperature increases.
In the cooler channel (AIA 171~\AA), the speed is about 120 km s$^{-1}$.
The wavenumber ($k$) is shorter than $\sim$66 Mm, and its frequency ($\omega$) is longer than 10 minutes.
In comparison to the cool channel, the wave speeds in hot channels are faster, and their wavelengths ($\sim$142~Mm) are longer.
The wave speeds in the AIA 193 and 211~\AA\ channels are $\sim$162 km s$^{-1}$ and $\sim$211 km s$^{-1}$, respectively.
The quasi-stationary features that exist at zero $\omega$ and $k_{r}$ in AIAC disappear in AIAH.
Our method contains the measurement errors determined from 10 estimations of the speeds for the different channels.
The estimated mean speeds are $\sim$123 km s$^{-1}$, $\sim$172 km s$^{-1}$, and $\sim$232 km s$^{-1}$, and
their standard deviation errors are $\pm$3 km s$^{-1}$, $\pm$6 km s$^{-1}$, and $\pm$24 km s$^{-1}$ for 171, 193, and 211 A, respectively.

We classified the plumes into two groups: active (L1, C, R1) and quiet plumes (L2, R2).
We added the Fourier power for the active plumes (group I: L1, C, R1) and the quiet plumes (group II: L2 and R2), and investigated their dispersion relations in the  AIA 171~\AA\ and AIA 193~\AA\ channels.
Figure 7 shows the relationship between the two groups in  the low and high AIA temperature channels, AIAC and AIAH.
In AIAC, the wavenumber and frequency of both groups have a similar trend,
but the speed of group I ($\sim$130 km s$^{-1}$) is higher} than that of group II ($\sim$116 km s$^{-1}$).
However, the wavenumber and frequency in AIA 193~\AA\ are different for groups I and II.
The power of AIA 193~\AA\ of group I is concentrated at low frequency ($\sim$30 min) and low wavenumber, while that of group II has a broad range of frequencies and wavenumbers as like AIA 171~\AA\ in group I.
In AIAH, the speed ($\sim$185 km s$^{-1}$) of group I is higher than the speed of group II  ($\sim$152 km s$^{-1}$).

PIDs are detected not only in plumes, but also in interplume regions \citep[e.g.,][]{2010ApJ...718...11G}. Thus, we also checked whether PIDs in the interplume regions have speed temperature dependency by applying the same method as for PIDs in plume regions.
As shown in Figure 8, we took three interplume strips as far as possible from the plumes, and then inspected the PIDs and derived their Fourier powers for the different channels.
As a result, we cannot identify any clear signature of speed dependency on the temperature from PIDs in the interplume regions.
This may due to the low signal to noise ratio because of the low intensity in the interplume regions, especially in the AIAH channels.
Other possibilities are thermal misbalance \citep[e.g.,][]{2019PhPl...26h2113Z} and steady upflows which is a property of the equilibrium \citep[e.g.,][]{2020ApJ...900L..19C} in the interplume regions, which could affect the apparent phase speed of slow waves. It is also noteworthy that the slow waves seen in the structures of different temperature propagate along different paths in the interplume regions.

\section{Summary and Discussion} \label{sec:summary}

Five plumes in the polar coronal hole in the northern hemisphere observed during the TSE 2017 have been investigated using the LASCO C2, SDO/AIA, Hinode/XRT, and DICE data.
The plumes were traced back from the high corona ($\sim$5.5 R$_\odot$ in LASCO C2 FOV) to the chromosphere (AIA 304~\AA), and the footpoints of the plumes were investigated.
We applied 2D Fourier transformation to the space-time plot of the plumes in the AIA 171, 193, and 211~\AA\ channels, and obtained an empirical dispersion relation of PIDs in the plumes.
From the dispersion relation, we found that in all analysed plumes, PIDs show a temperature dependence of their speed (i.e., AIAH propagation speed is larger than that of AIAC one).
This tendency is more evident when we added the Fourier power of all plumes in the different channels.
AIAC has broader ranges of frequencies ($\omega$) and wavenumbers ($k$) than those of AIAH.
The non-moving features located at zero-frequency and zero-wave numbers in AIAC disappear in AIAH,
in which high power signals are located in the lower frequency and wavenumber.

The quasi-periodic intensity change in Figure 4 is similar to the reported by Tian et al. (2011) who concluded that the intensity change represents outflow (jets)
because they did not find an obvious increasing trend in the speed with the increasing temperature.
Pucci et al. (2014) reported that the perturbations in AIAC have a higher speed than those in AIAH.
They concluded that the radiation variation in the plumes represents outflow material since their events did not show the temperature dependence which is expected from the magnetoacoustic wave hypothesis.
In this study, we have found a temperature dependence of the speed by applying the Fourier transformation.
In addition, we found that the observed ratio ($\sim$1.3) of the speeds in the AIA 171 and 193~\AA\ channels are similar to the expected theoretical values  \citep[1.25,][]{2006RSPTA.364..473N, 2014ApJ...793...86P}.
These results could be an evidence that supports the wave hypothesis.

The events studied here may differ in their properties from those of \citet{2014ApJ...793...86P},
who studied cooler plumes \citep[see, Figure 1 in][]{2014ApJ...793...86P}.
Here, we have investigated two groups of plumes that have different properties. By measuring temperatures from the XRT and DICE observations, we divided plumes depending on the existence of the high-temperature CBPs.
Active plumes are bright and associated with jets and macrospicules ejected from high-temperature CBPs.
These plumes extend as brighter (denser) structure into the higher corona ($\sim$5.5 R$_\odot$).
The quiet plumes are less bright, diffuse, and quasi-stable structures and possibly represent the decay stage of the plume evolution. Propagation speeds of PIDs in active plumes are higher than those of the quiet plumes, and this tendency is more significant in AIAH. The active and quiet plumes have different wavenumbers and frequency patterns in AIAH, but their difference is not discernible in AIAC.
It is noteworthy mentioning that the active plumes associated with CBPs may be an early stage of the plume's evolution  \citep{[see][]2003A&A...398..743D}.
The disturbance speed in AIAC is lower than that of AIAH, but the difference is not more significant than active plumes (group I).
From the AIAH movie from August 19 to 23, it is noted that the CBP in the foopoints of quiet plume L2 disappeared and the plume became quieter by the time of the eclipse observations on August 21.
The footpoint behavior of R2 could not be inspected because the footpoint of R2 plume moves behind the limb.

Our results lead us to conclude that the emission intensity variations in the studied plumes represent slow magnetio-acoustic waves. However, we cannot rule out the co-existence of plasma outflows with slow waves. The plumes' flows are likely to consist of cool and dense material from macrospicules and high temperature jets from CBPs. The flows seen in the low temperature channels may be plasma flows produced by quasi-static upward evaporation heated by thermal conduction during magnetic reconnection at the boundary of the supergranulation cell \citep{1995ApJ...452..457W}.

Due to the overlapping of plume structure with fore- and background, and the lack of the observations at higher resolution, the triggering mechanism of the polar plume is still not well understood.
Together with the new opportunity with next generation coronagraph, CODEX (Coronal Diagonostic Experiment) that provides electron temperature and speed structure of plumes \citep{2017JKAS...50..139C},
imaging and spectroscopic observations using high-resolution telescopes from the ground (e.g., Goode Solar Telescope (GST) in Big Bear Solar Observatory) and space (e.g., Interface Region Imaging Spectrograph (IRIS))
focusing on the plumes' roots in polar coronal holes is essential to address a new model for understanding the physical mechanisms that take place in active plumes.


\acknowledgments
This work was supported the Korea Astronomy and Space Science Institute under the R\&D program (2020-1-850-07).
IHC acknowledges a support from the National Research Foundation of Korea (NRF-2019R1C1C1006033).
MM was supported by the Brain Pool Program of the National Research Foundation of Korea (NRF-2019H1D3A2A01099143).
KSL was supported by the National Research Foundation of Korea (NRF-2020R1A2C2004616).

\newpage
\begin{figure}
\center
\includegraphics[scale=1.5, angle=0]{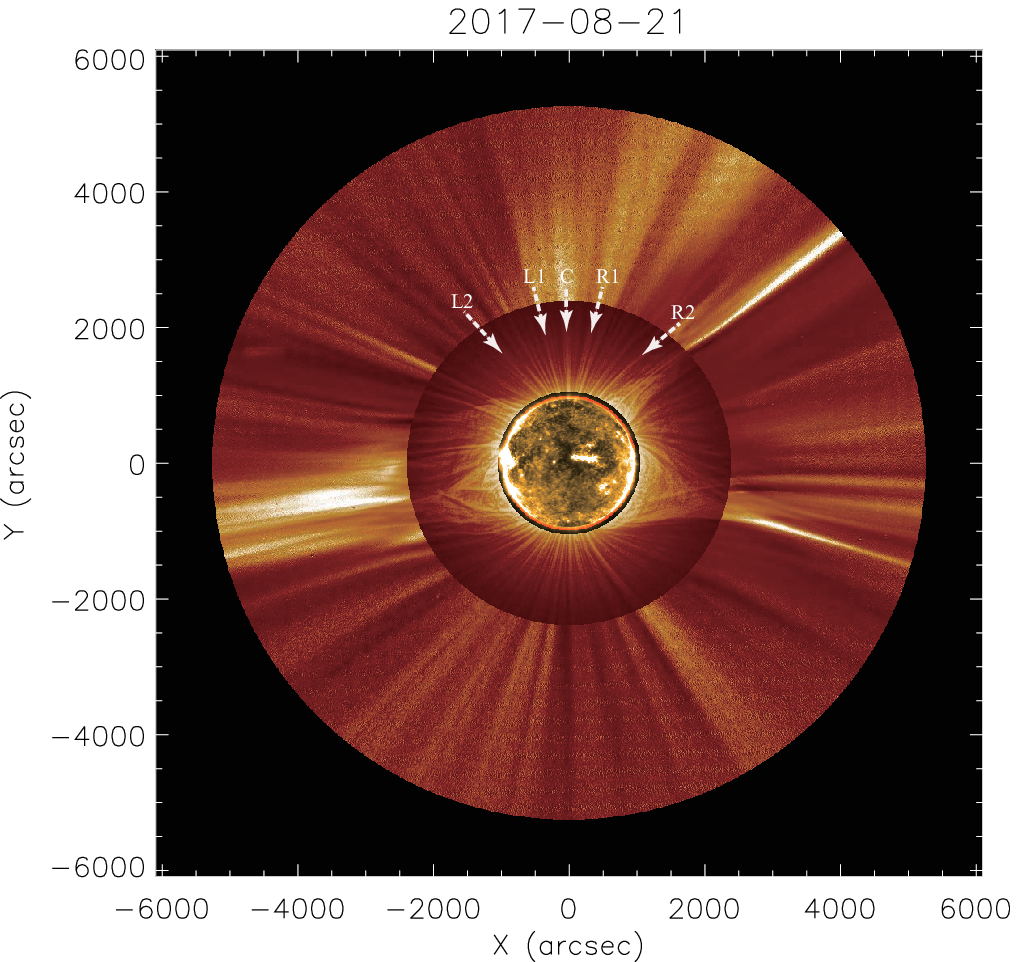}
\caption{
A composite image of LASCO C2 (R $>$ 2.5 R$_\odot$), White Light (WL) image (1.1 R$_\odot$ $\leq$ R $\leq$ 2.5 R$_\odot$), and AIA 171 \AA\ (R $<$ 1.10 R$_\odot$). The WL image is produced by using the WL images taken by Cannon 400 mm camera during the total eclipse of 2017 August 21 (17:35 UT - 17:37 UT), and the AIA 171 \AA\ was obtained at 17:45 UT. The LASCO C2 image is the mean intensity between 16:00 UT and 20:00 UT. Arrows denote the plumes in the north coronal Hole (CH) for this study.
}
\label{fig1}
\end{figure}

\begin{figure}
\center
\includegraphics[scale=1.0]{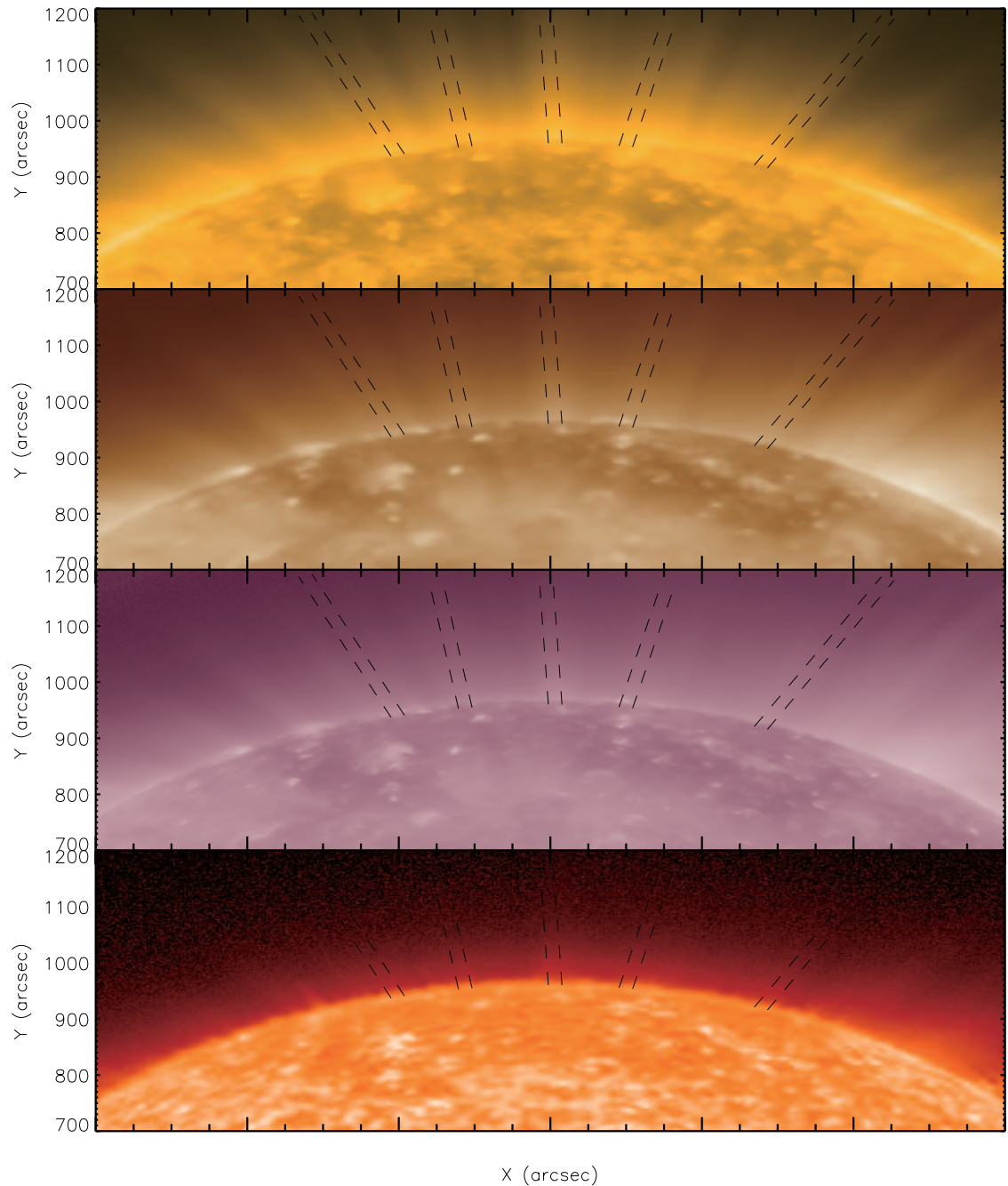}
\caption{
Averaged SDO/AIA images between 16:00 UT and 17:40 UT to enhance the structure of plumes denoted by arrows in Figure 1. The AIA 171, 193, 211, and 304 \AA\ are plotted in the panels from the top to the bottom. Dotted slots along the plumes are used to inspect activities at the bottom of the plumes and to trace flows in the plumes marked with arrows in Figure 1. See the accompanying animation from 16:00 UT to 17:40 UT, which shows outflows and jets in the plumes, and recurrent spicules and macro-spicules in the base of the plumes. An animation of this figure is available in the online journal.
}
\label{fig2}
\end{figure}

\begin{figure}
\center
\includegraphics[scale=1.3]{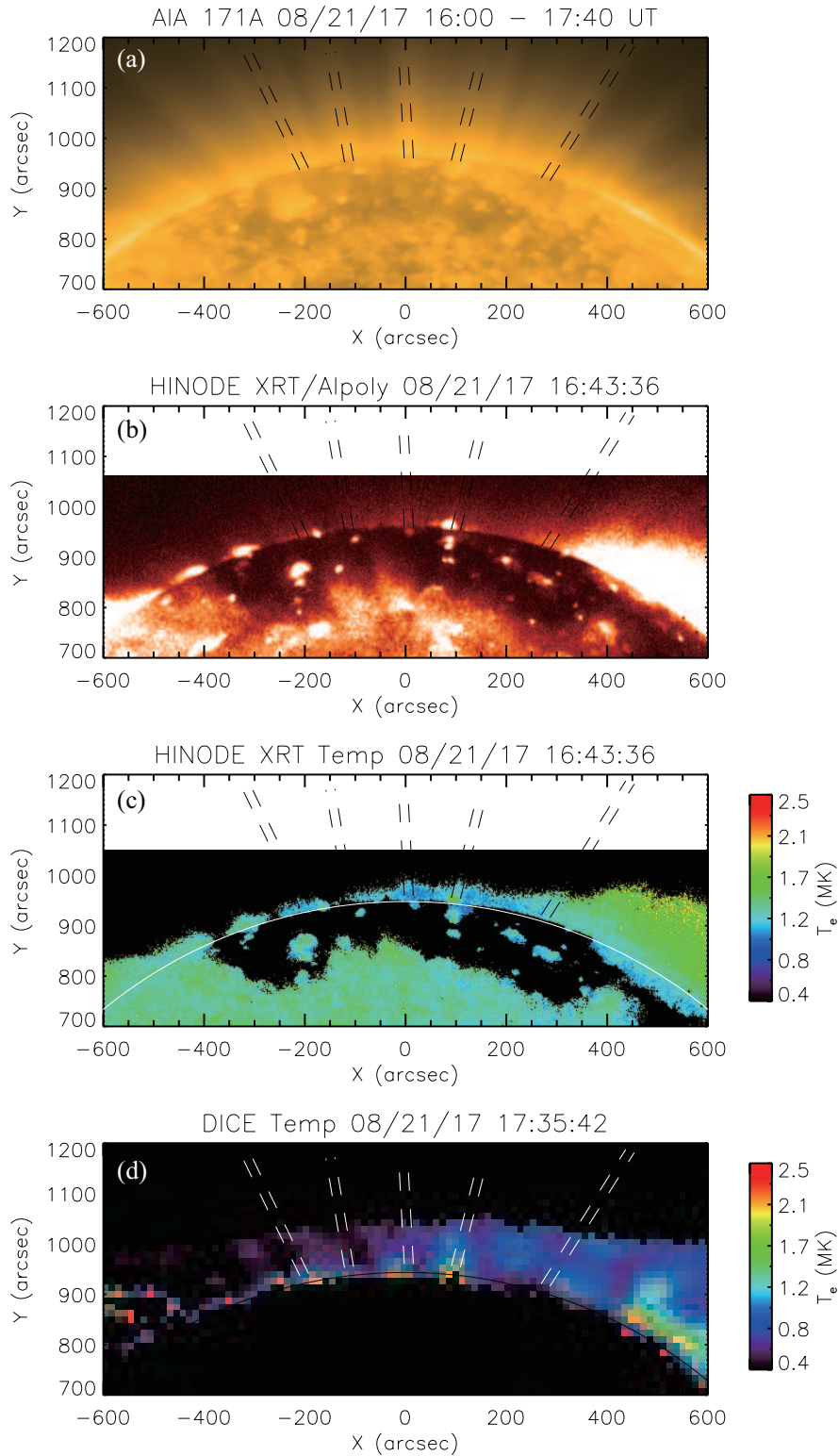}
\caption{
EUV and X-ray images together with coronal temperature maps from Hinode/XRT and 2017 total solar eclipse experiment, DICE. (a) AIA 171 \AA\ image taken 17:36 UT, (b) Hinode XRT/Al-poly image before the eclipse, (c) temperature map calculated by the ratio method using Al-mesh and Al-poly data, (d) temperature map calculated by the ratio method using polarized brightness images taken 3934 \AA\ and 4025 \AA. The dashed slot denote the same locations in Figure 2.
}
\label{fig3}
\end{figure}

\newpage
\begin{figure}
\center
\includegraphics[scale=1.0, angle=0]{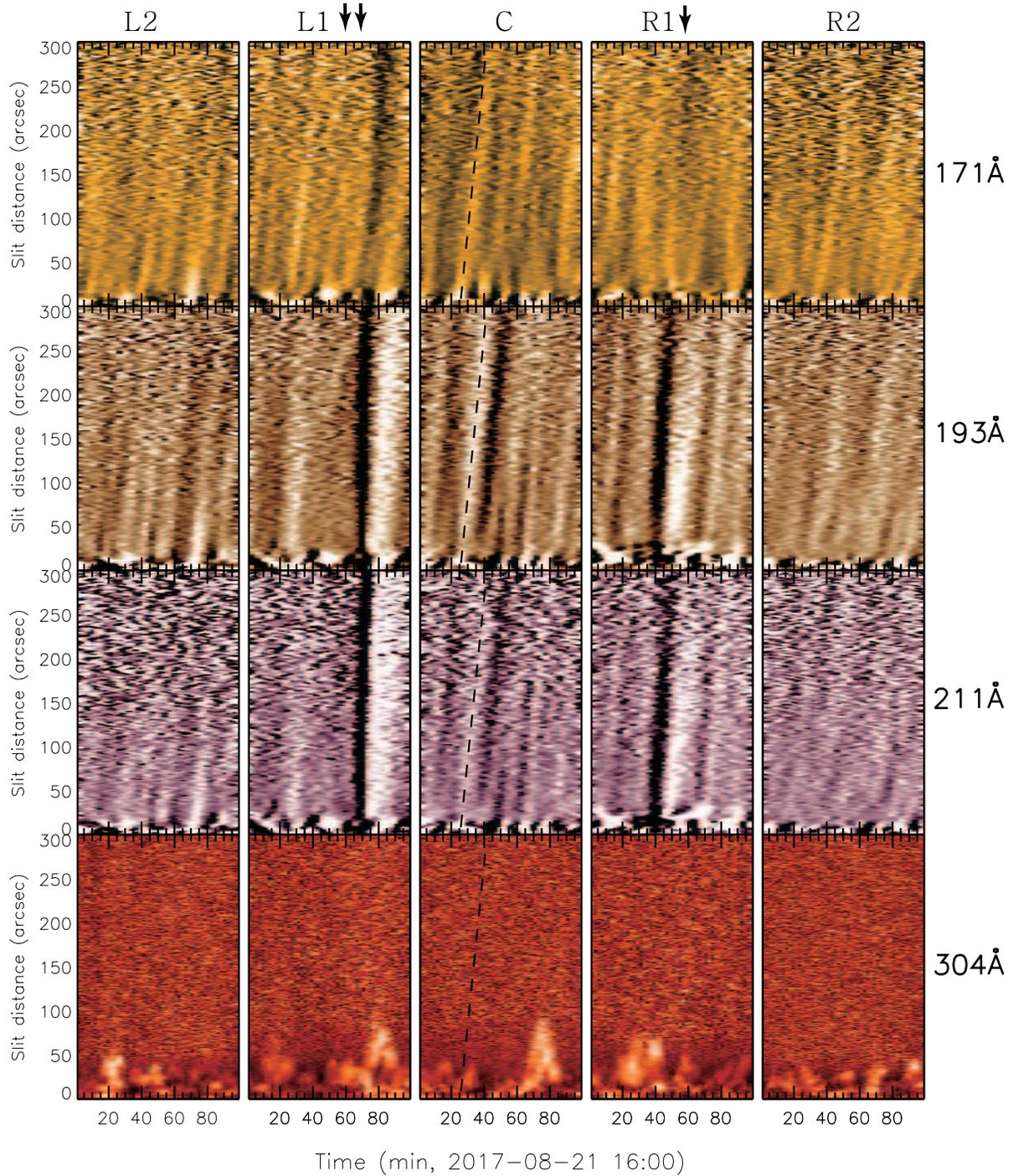}
\caption{
Space-time plots of the de-trended AIA 171, 193, 211, 304 \AA\ images from top to bottom. From the left to the right panels, the space-time plots are displayed for the regions of L2, L1, C, R1, R2 in Figure 3(b). Dashed line denotes an example of strips, and arrows indicate AIAC strips are missing in AIAH, and vice versa (see text).
}
\label{fig4}
\end{figure}

\begin{figure}
\center
\includegraphics[width=1.0\textwidth]{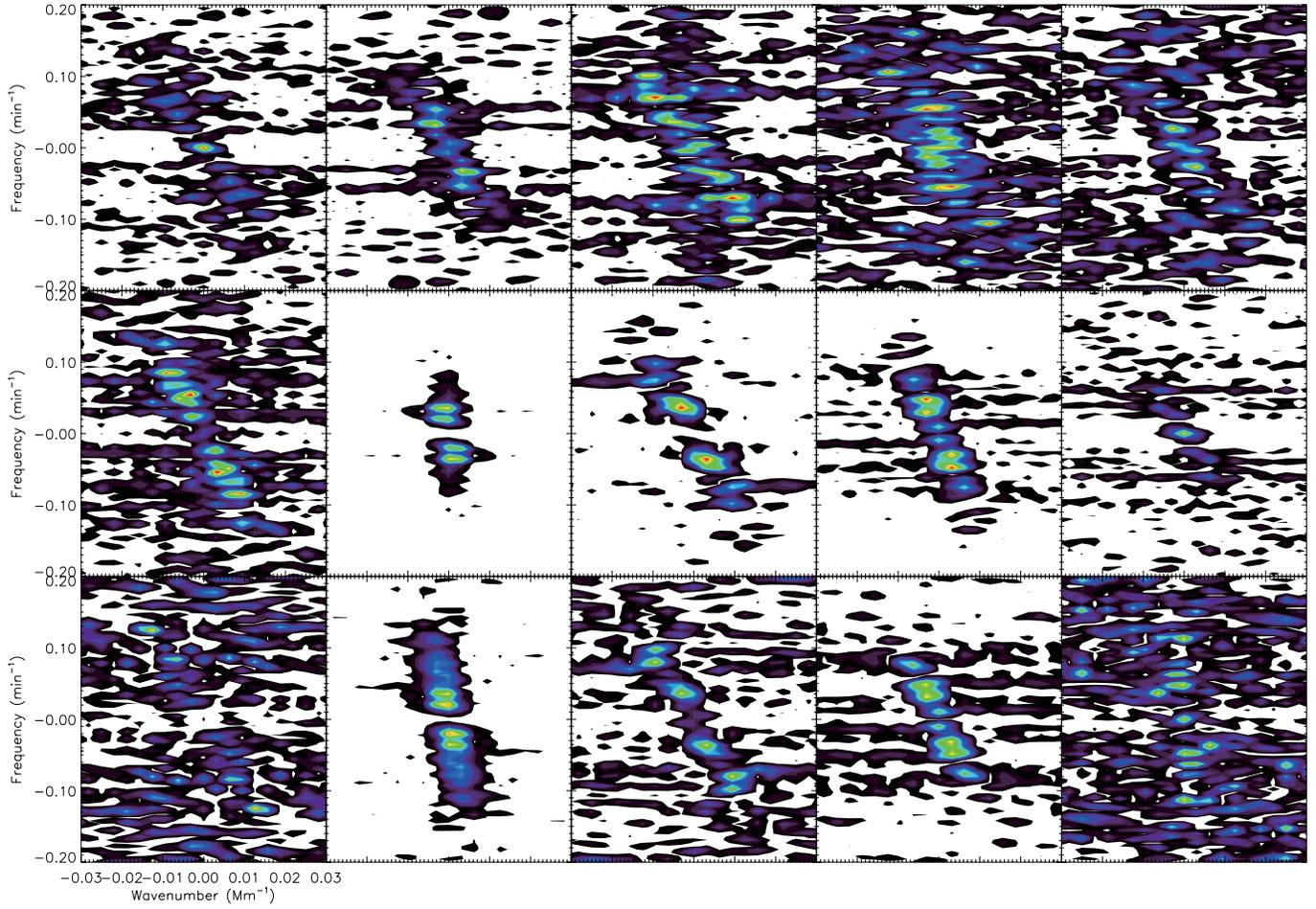}
\caption{
Wave number - frequency ($k - \omega $) maps for the five plumes (from the left to the right) and three AIA channels of 171, 193, and 211 \AA\ (from the top to the bottom).
}
\label{fig5}
\end{figure}

\newpage
\begin{figure}
\center
\includegraphics[width=0.8\textwidth]{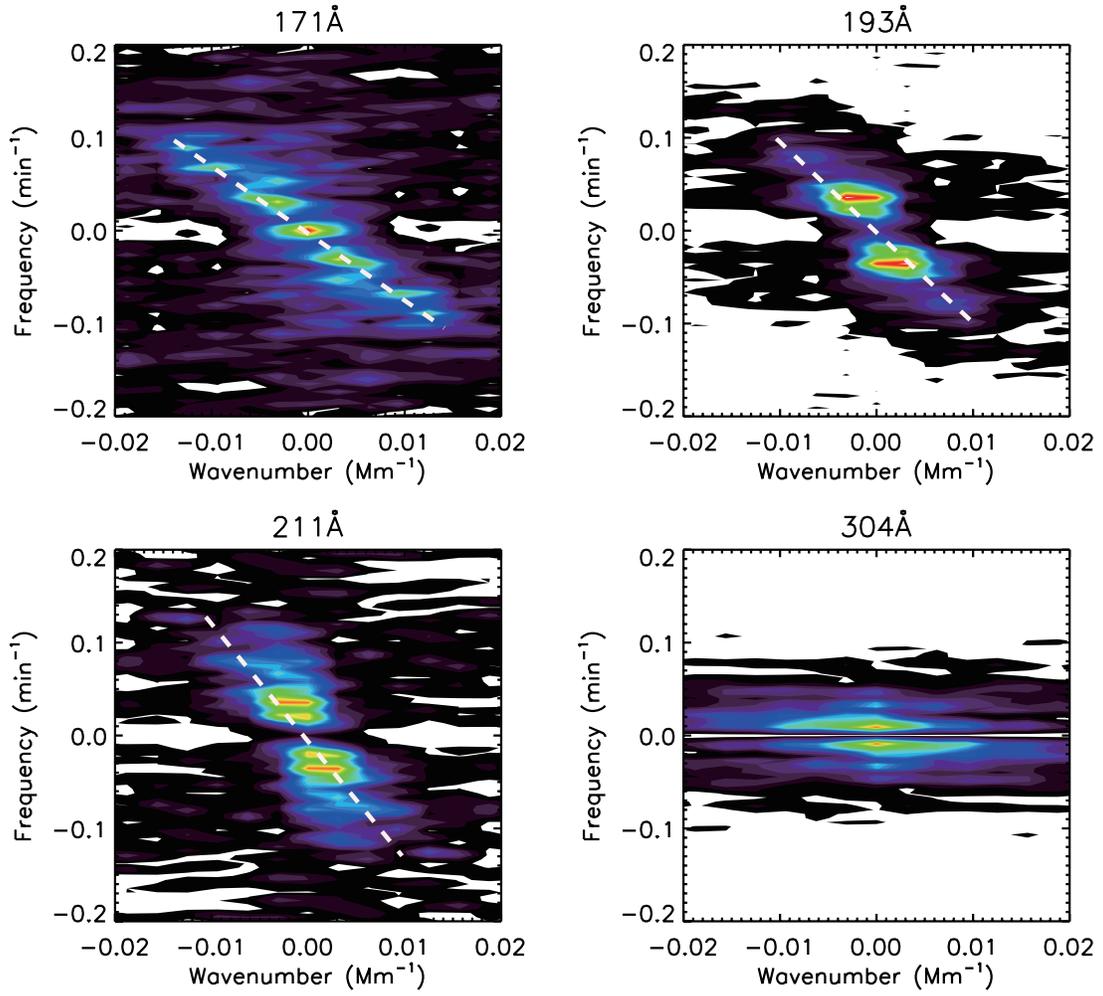}
\caption{
Averaged wave number - frequency ($k - \omega $) maps for all plumes in the AIA 171, 193, 211, and 304 \AA\ channels for both groups I and II.
The slop of the diagonal dashed line is used to obtain the propagation speed.
}
\label{fig6}
\end{figure}

\newpage
\begin{figure}
\center
\includegraphics[width=0.8\textwidth]{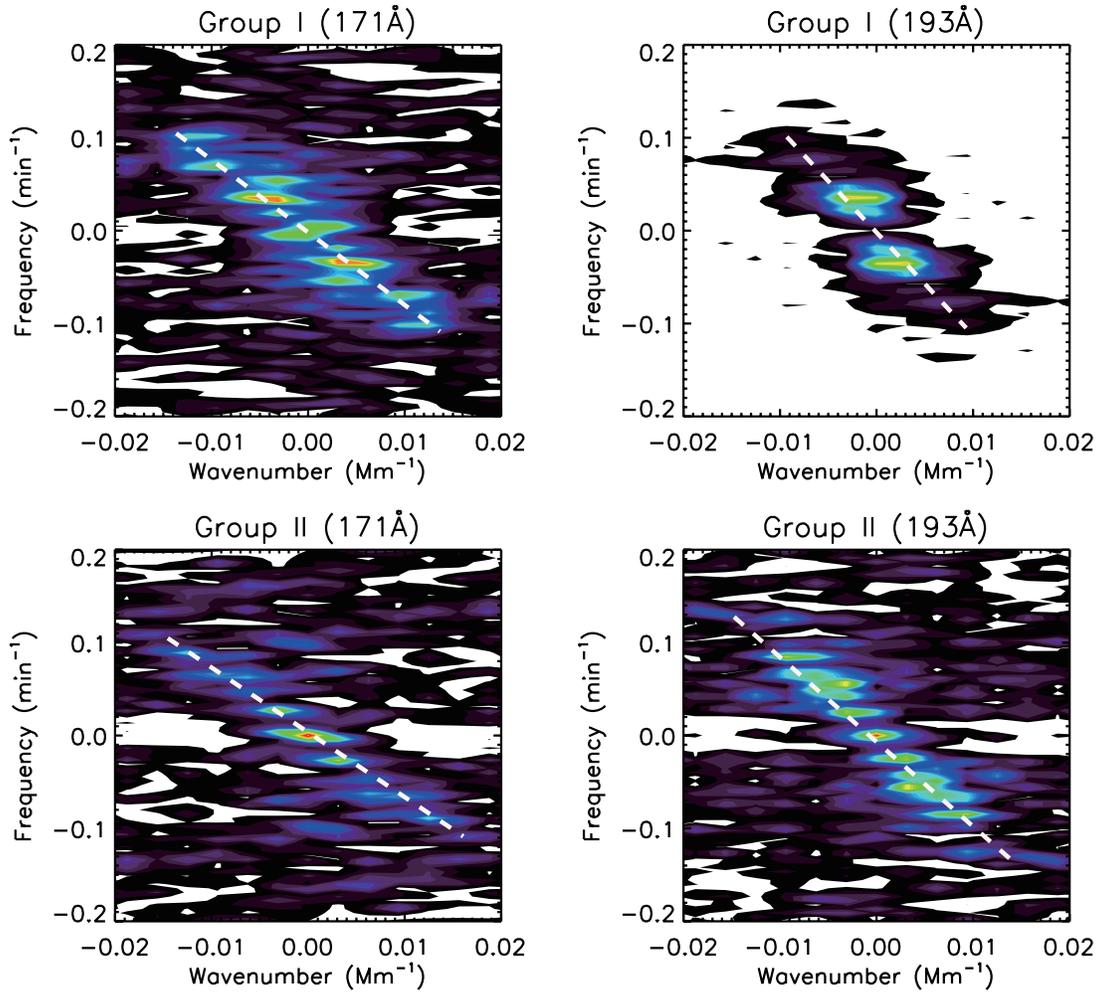}
\caption{
Averaged wave number - frequency ($k - \omega $) maps of AIA 171 and 193 \AA\ for plumes with active foot-points (upper panels) and quiet foot-points (lower panels).
The slop of the diagonal dashed line is used to obtain a propagation speed.}
\label{fig7}
\end{figure}

\newpage
\begin{figure}
\center
\includegraphics[width=1.0\textwidth]{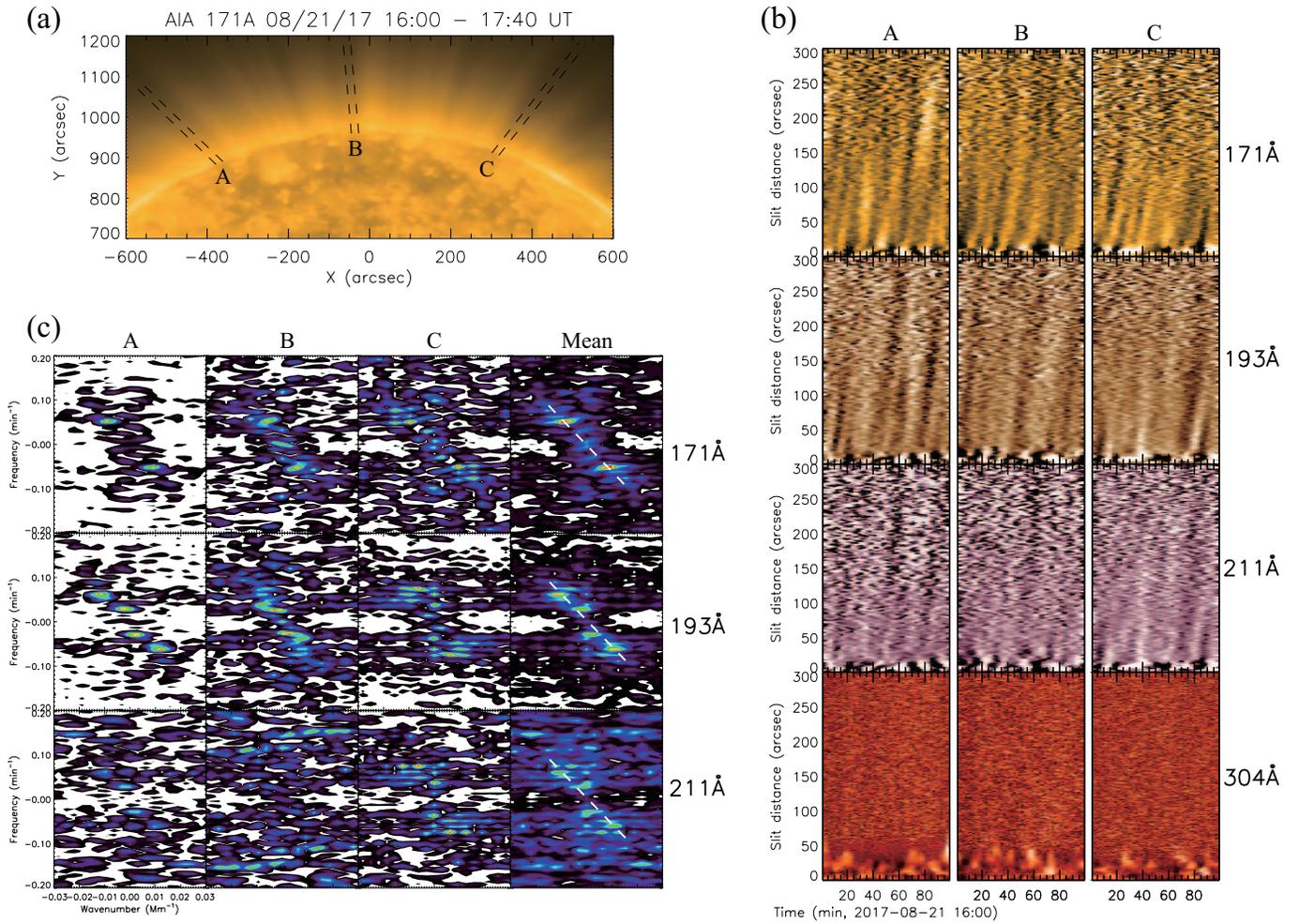}
\caption{
Locations of three interplumes (`A', `B', `C') and their PIDs properties. (a) AIA 171 \AA\ image is denoted with interplumes' locations, (b) space-time plots of the AIA 171, 193, 211, and 304 \AA\ images from top to bottom, and (c) wave number - frequency ($k - \omega $) maps of AIA 171, 193 and 211 \AA\ for the interplumes. The last column of the wave number - frequency maps is the mean power map of the all interplumes.}
\label{fig7}
\end{figure}

\clearpage
\newpage
\begin{table}[ht]
\center
\caption{Activities of the plumes}
\begin{tabular}{cccccc}
\hline
      & L2  & L1  & C & R1 & R2 \\
\hline
Jets occurrence    & no & yes & yes & yes & no \\
Macro spicule     & quiet & active & active & active & quiet \\
Foot-point Temp & low & high & high & high & low \\
Plume activity  & quiet & active & active & active & quiet \\
\hline
\label{table1}
\end{tabular}
\end{table}

\end{document}